\documentclass[aps,pra,reprint,superscriptaddress,floatfix,longbibliography]{revtex4-2}

\usepackage{amssymb}
\usepackage{graphicx}   
\usepackage{color}      
\usepackage{hyperref}   
\usepackage{dcolumn}

\def\lsco{La$_{2-x}$Sr$_x$CuO$_4$}
\def\lbco{La$_{2-x}$Ba$_x$CuO$_4$}

\def\lnsco{La$_{1.6-x}$Nd$_{0.4}$Sr$_x$CuO$_4$}

\begin{document}

\title{Testing for pair-density-wave order in La$_{1.875}$Ba$_{0.125}$CuO$_4$}

\author{P. M. Lozano}
\affiliation{Condensed Matter Physics and Materials Science Division, Brookhaven National Laboratory, Upton, New York 11973, USA}
\affiliation{Department of Physics and Astronomy, Stony Brook University, Stony Brook, New York 11794-3800, USA}
\author{Tianhao Ren}
\author{G. D. Gu}
\author{A. M. Tsvelik}
\author{J. M. Tranquada}
\email{jtran@bnl.gov}
\affiliation{Condensed Matter Physics and Materials Science Division, Brookhaven National Laboratory, Upton, New York 11973, USA}
\author{Qiang Li}
\email{liqiang@bnl.gov}
\affiliation{Condensed Matter Physics and Materials Science Division, Brookhaven National Laboratory, Upton, New York 11973, USA}
\affiliation{Department of Physics and Astronomy, Stony Brook University, Stony Brook, New York 11794-3800, USA}

\date{\today} 

\begin{abstract}
Charge order is commonly believed to compete with superconducting order.  An intertwined form of superconducting wave function, known as pair-density-wave (PDW) order, has been proposed; however, there has been no direct evidence, theoretical or experimental, that it forms the ground state of any cuprate superconductor.  As a test case, we consider \lbco\ with $x=1/8$, where charge and spin stripe orders within the CuO$_2$ planes compete with three-dimensional superconducting order.  We report measurements of the superconducting critical current perpendicular to the planes in the presence of an in-plane magnetic field.  The variation of the critical current with orientation of the field is inconsistent with a theoretical prediction specific to the PDW model.  It appears, instead, that the orientation dependence of the critical-current density might be determined by a minority phase of $d$-wave superconductivity that is present as a consequence of doped-charge inhomogeneity.
\end{abstract}

\maketitle

\section{Introduction}

In a superconductor, a collective state of paired electrons supports dissipationless transport, corresponding to current flow without resistance.  For a solid with charge order, there is a static spatial modulation of the density of conduction electrons.  While charge order has now been observed in most cuprate superconductors \cite{comi16,fran20}, charge and superconducting orders are typically viewed as competitors \cite{keim15,imad21,chan12a,capl15}.   An extreme case occurs in \lbco\ (LBCO) with $x=1/8$, where the crystal structure at low temperature has anisotropic Cu-O bonds that stabilize charge and spin stripe orders \cite{fuji04,huck11}.  Unusual two-dimensional (2D) superconductivity develops at the onset of spin-stripe order \cite{li07,frad15}.

\begin{figure*}[t]
 \centering
   \includegraphics[width=1.9\columnwidth]{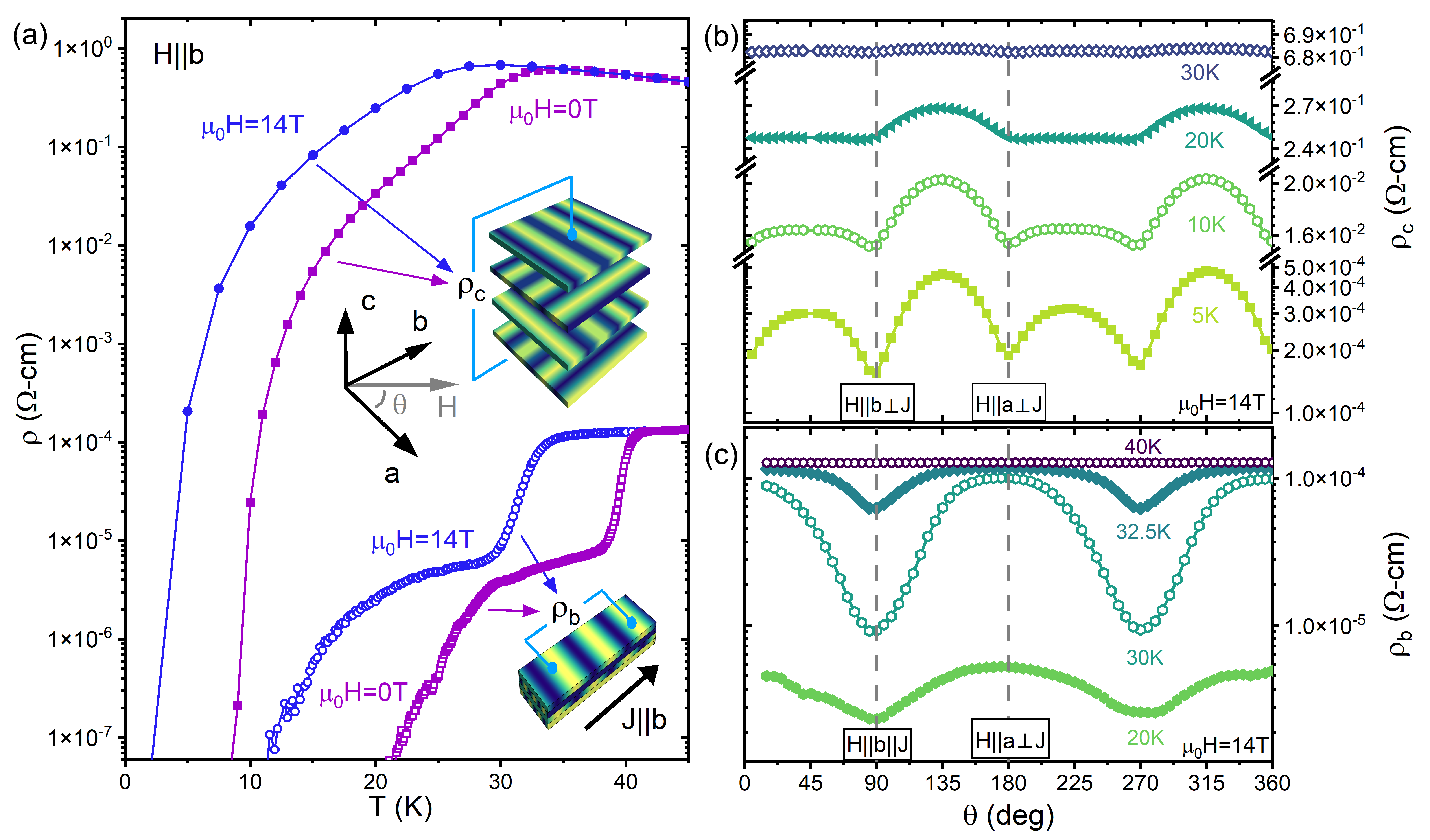}
    \caption{\label{fg:rho}  (a) Resistivity vs.\ temperature for $\rho_b$ (open symbols) and $\rho_c$ (filled symbols) in zero field (magenta squares) and full field of 14 T applied along the {\it b} axis (blue circles), corresponding to field angle $\theta=90^\circ$ measured relative to the $a$ axis. ($a$ and $b$ are equivalent, and correspond to Cu-O bond directions.) Upper inset indicates the proposed PDW order, with the superconducting wave function oscillating from positive (dark) to negative (light), and rotating by 90$^\circ$ between layers \protect\cite{berg07}.  Insets also indicate relative positions of voltage contacts.  (b) Variation of $\rho_c$ with $\theta$ at $T=5$, 10, 20, and 30~K.  (c) Variation of $\rho_b$ with $\theta$ at $T=20$, 30, 32.5, and 40~K.  }
\end{figure*}

Evidence for the 2D superconductivity is illustrated in Fig.~\ref{fg:rho}(a), where one can see that the in-plane resistivity, labelled $\rho_b$ (where $b$ is one of the two equivalent axes aligned with Cu-O bonds), in the absence of a magnetic field shows a substantial drop at $\sim40$~K, indicating the onset of phase-disordered superconductivity, with phase order developing below 20~K in the form of a Berezinskii-Kosterlitz-Thouless transition \cite{li07}.  Meanwhile, the resistivity along the $c$ axis, $\rho_c$ (measured perpendicular to the planes), remains large until the temperature drops below 10~K, demonstrating the 2D character of the superconducting fluctuations.  Application of a strong in-plane magnetic field lowers these transition temperatures, but they remain finite.

It is extremely unusual to observe 2D superconductivity within equivalent layers of a bulk crystal because one would normally expect some type of effective Josephson coupling between neighboring layers that results in three-dimensional superconducting order.  To explain the apparent frustration of the interlayer Josephson coupling \cite{li07,taji01}, PDW order was proposed \cite{hime02,berg07,agte20}.  In this state, the pair wave function has extrema on the charge stripes, where the amplitude oscillates from positive to negative on neighboring charge stripes.   The suggested PDW state corresponds to a situation where the superconducting pairs have finite momenta along the direction of the charge modulation.  The frustration of the interlayer coupling comes from a 90$^\circ$ rotation of the PDW order between layers, following the pinning of the charge stripes to the lattice anisotropy \cite{zimm98}, as indicated in the upper inset of Fig.~\ref{fg:rho}(a).

While the PDW proposal is consistent with experiment, its relevance remains uncertain. The PDW is a strongly-correlated state that is difficult to reconcile with the conventional theory of superconductivity \cite{bard57}, which is based on a model of nearly-free, spatially-extended electron waves.  On the other hand,  evaluations of relevant theoretical models appropriate to hole-doped cuprates using advanced numerical techniques find that, while there is evidence for charge- and spin-stripe orders for a hole concentration of 1/8, the measure of superconducting coherence is strongly depressed and spatially uniform \cite{whit09,whit15,jian21c}.  Calculations show that the PDW state is close in energy to other solutions \cite{corb14}, but none have identified conditions where it is the ground state.  

Yang \cite{yang13} proposed an experimental test directly sensitive to the putative PDW state in LBCO.  He noted that the mismatch between the momenta of the Cooper pairs located in adjacent CuO$_2$ planes can be reduced by application of an in-plane magnetic field \cite{yang00b}; 
measurements on the closely-related compound La$_{1.7}$Eu$_{0.2}$Sr$_{0.1}$CuO$_4$ have demonstrated that a strong in-plane field can reduce $\rho_c$ \cite{shi20b}.  A phase-sensitive prediction is that the superconducting critical-current density along the $c$ axis should be maximum when the field is at 45$^\circ$ to the Cu-O bonds.  Unfortunately, our results find the maxima to occur when the field is parallel to Cu-O bonds, as previously observed in stripe-ordered \lnsco\ \cite{xian09}.  It now appears that the anisotropy might be the result of an ``extrinsic'' effect due to inevitable charge inhomogeneity \cite{ren22}.

The rest of this article is organized as follows.  After a brief description of the experimental methods, the results are presented in Sec.~III.  A comparison with previous results and a discussion of a new proposed interpretation are given in Sec.~IV.  Our conclusions appear in Sec.~V.

\section{Experimental Methods}

Single crystals of LBCO with $x = 1/8$ studied here were grown in an infrared image furnace by the floating-zone technique. They are pieces from the same cylindrical crystal used previously to characterize two-dimensional fluctuating superconductivity \cite{li07}. Single-crystal samples were cut and aligned into slabs, then fixed on a 0.5-mm-thick sapphire substrate. The imperfection in the sample alignment, estimated from X-ray diffraction, is less than 0.5$^\circ$. 
For transport measurements, current contacts were made at the ends of the longest dimension of crystals ({\it e.g}., a $c$-axis-oriented crystal with dimensions along axes $c \times b \times a$ of  $3.50 \times 0.94 \times 0.20$~mm$^3$) to ensure uniform current flow, while the voltage contacts were made on both the top and side of the crystals. We used a low-temperature contact annealing procedure \cite{li07} leading to low contact resistance ($< 0.2$~$\Omega$) that allows us to measure the resistivity over seven orders of magnitude.  The angle-dependent magnetoresistance (ADMR) was measured using the 4-point probe in-line method in a Quantum Design Physical Property Measurement System (PPMS) equipped with a 14-T superconducting magnet. The resistivity measurements have been performed with the current applied along either the $a(b)$-direction or the $c$-direction using dc and ac transport options with a current range of 50~$\mu$A -- 1~mA. Both dc and ac methods produced the same results. The data shown are from the ac transport measurements (17 Hz). For crystal alignment with magnetic field, horizontal and vertical sample rotators were used with the angular resolution $\sim 0.1^\circ$. Temperature dependent ADMR data were taken from 1.8 to 300~K, at various fields up to 14~T.  ADMR data at fixed temperatures and magnetic fields were taken in-situ with a vertical sample rotator as a function of the in-plane magnetic field angles ($\theta$) in a range of $-15^\circ$ to 360$^\circ$.  The ADMR results were confirmed by measurements on a second crystal \footnote{The measurements on the second crystal were done after we had prepared a manuscript on the results from the first crystal. The second set of measurements led us to realize that the $a$ and $b$ axes of the first crystal had been misidentified, as confirmed by analysis of X-ray Laue diffraction}.

Note that we use $a$ and $b$ to label the in-plane crystal axes aligned with the Cu-O bonds, which are equivalent and indistinguishable in the low-temperature-tetragonal phase \cite{huck11} relevant to all measurements presented here.

\begin{figure*}[t]
 \centering
  \includegraphics[width=1.35\columnwidth]{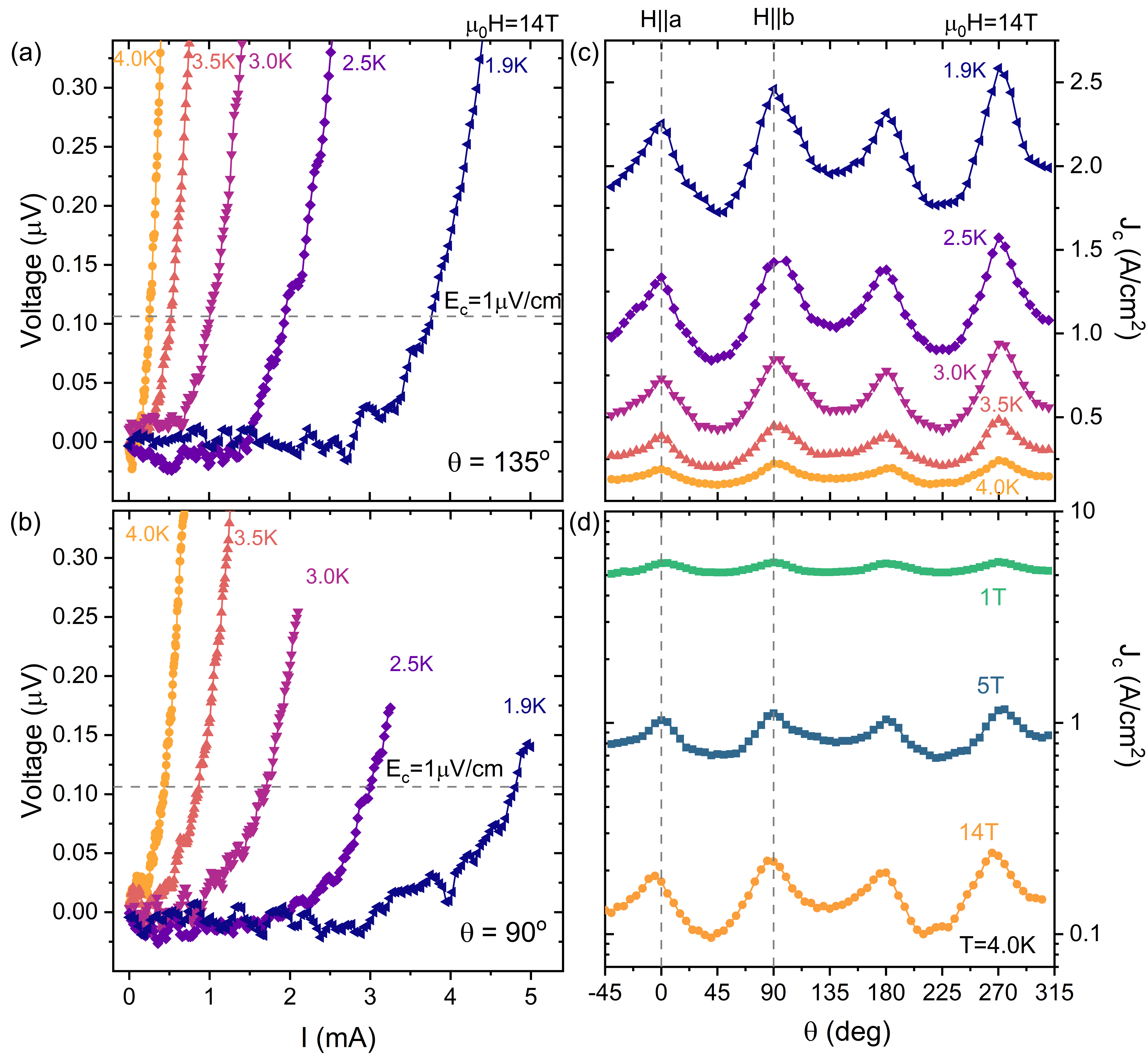}
    \caption{\label{fg:ccd}   Examples of voltage vs.\ current applied along the $c$ axis with the in-plane magnetic field of 14 T oriented (a) at 45$^\circ$ to the $b$ axis ($\theta = 135^\circ$), and (b) along the $b$ axis ($\theta=90^\circ$), for several temperatures.  Dashed line indicates the threshold criterion, corresponding to an electric field $E_c=1$~$\mu$V/cm, used to determine the critical current.  (c) Variation of the critical current density along $c$, $J_c$, with field angle $\theta$ for several temperatures and a magnetic field of 14 T.  Maxima are distinctly aligned with the directions with the Cu-O bond directions.  (d) Similar to (c), but comparison of results for three values of the magnetic field (1, 5, and 14 T) at $T=4$~K; note that the scale for $J_c$ is logarithmic.}
\end{figure*}

\section{Results}

Consider the ADMR results in Fig.~\ref{fg:rho}(b), obtained in the maximum field of 14~T.  There is no significant modulation at $T=30$~K, where, as one can see in Fig.~\ref{fg:rho}(a), $\rho_c$ is at its maximum; however, oscillations become apparent at 20 K, where $\rho_c$ has begun to decrease, and they become stronger with further cooling.  The minima in $\rho_c$ occur whenever the field is along a Cu-O bond direction.

The lack of perfect 4-fold symmetry is a consequence of the sample shape.  The crystal for this measurement is longest along $c$, and it has unequal widths of the $a$ and $b$ faces, as described in the previous section.  This leads to anisotropy in the demagnetization factor \cite{proz18}, which means that the internal magnetic field is not precisely identical when the field is along $a$ or $b$.

For comparison, we show the impact of field orientation on in-plane resistivity $\rho_b$ in Fig.~\ref{fg:rho}(c).  In this geometry, we have an anisotropy that is controlled by the orientation of the field relative to the measurement current, which is along $b$, resulting from the variation in the Lorentz force on magnetic vortices.  The resistivity is a minimum when the current is parallel to the applied field, which means that we have a two-fold variation, and not the four-fold modulation of $\rho_c$, when the temperature is below the onset of in-plane superconductivity.

To explore the critical-current density along the $c$ axis, we have to cool to below 5~K.  Figures \ref{fg:ccd}(a) and (b) show examples of voltage vs.\ current measurements for field at $45^\circ$ to the $b$ axis and along the $b$ axis, respectively, and temperatures from 4 K down to 1.9 K.  Following standard procedure, we identify the critical current as the value at which the voltage crosses a threshold value indicated by the dashed line, which corresponds to an electric field along the $c$ axis of 1~$\mu$V/cm.  

The variation of the $c$-axis critical-current density, $J_c$, with field angle is plotted in Fig.~\ref{fg:ccd}(c) at maximum field for several temperatures.  As one can see, it peaks periodically when the field is along a Cu-O bond.  Figure \ref{fg:ccd}(d) shows that the effect is detectable with magnetic fields of smaller magnitude, as well.  Of course, at fixed temperature there is a large change in $J_c$ with field magnitude due to its effect on the superconductivity in the CuO$_2$ planes, as one can see in Fig.~\ref{fg:rho}(a).  The observed angle dependence of $J_c$ is precisely out of phase with the prediction based on orthogonally-stacked PDW order \cite{yang13}.

\section{Discussion}

The ADMR that we observe in $\rho_c$ below 30~K has the same fourfold symmetry and orientation as that reported for stripe-ordered \lnsco\ with $x=0.15$ \cite{xian09}.  In that work, it was attributed to anisotropic pinning of magnetic vortices by charge stripes.  That explanation seems unlikely given the fact that the ADMR is observed at temperatures where $\rho_b$ is finite, and without superconducting phase order within the planes there cannot be pinning of vortices.  We also note that the ADMR observed here is distinct from the normal-state ADMR reported for strongly overdoped cuprates such as \lnsco\ with $x=0.24$ \cite{fang22} and Tl$_2$Ba$_2$CuO$_{6+\delta}$ \cite{huss03}.

To evaluate an alternative explanation of the modulation in $\rho_c$ and $J_c$, it is necessary to take account of all possible superconducting paths in the sample.  Clearly, the decrease of $\rho_c$ below $\sim25$~K is not what one would expect from an ideal system of 2D superconducting layers with a uniformly-frustrated interlayer Josephson coupling.  The observed $T$-dependence of $\rho_c$ resembles the behavior of one-dimensional (1D) superconducting nanowires in which phase slips result in finite resistivity \cite{roga03}. It suggests that in the temperature interval between 4 and $\sim25$~K we have the peculiar situation of two types of liquids of superconducting pairs: one type involving 2D PDW order, and the other type consisting of pairs located on effective nanowires traversing the sample along the $c$ direction.  It is important to note that the effective 1D superconducting fluctuations along $c$ must be decoupled from the 2D PDW superconductivity.  If they were coherent with one another, this would provide an interlayer coupling between the PDW order in the layers and the superconductivity would immediately become 3D.  One likely origin of such a situation lies in charge inhomogeneities.  As discussed in detail elsewhere \cite{tran21a}, charge disorder is significant in cuprates, as demonstrated by local probes such as nuclear magnetic resonance \cite{sing02a}.  Hence, we can expect to have some patches in each plane with a local hole concentration $\gtrsim0.14$ that can support spatially-uniform superconductivity. Some of these patches will be able to couple along the $c$ axis, causing $\rho_c$ to drop.    A subset of these may form effective 1D ``trails'' crossing the sample.  Another contribution may come from crystallographic twin boundaries, where the local variation in symmetry \cite{zhu94} might allow finite patches of uniform superconductivity that could communicate along the $c$ axis.

The fraction of the full Meissner response observed at 2~K in a field of 0.2 mT parallel to the planes is only 0.1\%\ \cite{tran08}, which contrasts with a value of at least 20\%\ measured in polycrystalline \lsco\ for a large range of $x$ \cite{naga93}.  This is compatible with a minority phase of uniform superconductivity being responsible for the drop of $\rho_c$ to zero at low temperature.  A mechanism explaining the observed ADMR in terms of such a minority phase has been proposed in \cite{ren22}.  Assuming that PDW order is present, as suggested by the high-temperature onset of 2D superconductivity, the lack of a dominant response to an in-plane magnetic field might be evidence for the strong degree of frustration of the interlayer Josephson coupling.  We should also note that positive phase-sensitive evidence for PDW order has been reported in studies of Josephson junctions with \lbco\ $x=1/8$ crystals \cite{hami18}.

There have been several previous reports of {\it local} PDW order by scanning tunneling microscopy.  These include detecting PDW order in the vicinity of magnetic vortex cores through interference with uniform superconductivity \cite{edki19} and through local periodic modulations of the superconducting gap \cite{du20}.  It is possible that local perturbations, such as a magnetic vortex core, may change the energy balance, stabilizing PDW locally even when the energetically-favored order in the bulk is spatially-uniform superconductivity \cite{loza21a}.  

\section{Conclusion}

In conclusion, we have observed a fourfold modulation of $J_c$ as a function of the orientation of an in-plane magnetic field.  The maximum $J_c$ occurs when the field is along a Cu-O bond direction, which is inconsistent with a prediction based on the idea of partial relief of the frustration of interlayer Josephson coupling due to PDW order \cite{yang13}.  The observations are consistent with an earlier study of stripe-ordered \lnsco\ with $x=0.15$ \cite{xian09}.  It appears that the anisotropy may actually be a consequence of minority regions of uniform superconductivity, as proposed in \cite{ren22}.

\section{Acknowledgments}
We acknowledge helpful comments from E. H. Fradkin, S. A. Kivelson, V. Oganesyan, and K. Yang.   This work was supported by the U.S. Department of Energy, Office of Basic Energy Sciences, Division of Materials Sciences and Engineering, under Contract No. DE-SC0012704.

\bibliography{LNO,theory}

\end{document}